\documentclass[aps,prd,nofootinbib,showkeys,floatfix,twocolumn,bibnotes,preprintnumbers]{revtex4-1}
\usepackage{graphicx,amsmath,amssymb,hyperref,natbib,slashed}
\usepackage[dvipsnames]{xcolor}
\usepackage{xspace}
\usepackage{bbm} 
\usepackage{lipsum}
\usepackage{subfigure}

\usepackage[utf8]{inputenc}

\newcommand*\diff{\mathop{}\!\mathrm{d}}

\newcommand{\be}{\begin{equation}}
\newcommand{\ee}{\end{equation}}
\newcommand{\bea}{\begin{eqnarray}}
\newcommand{\eea}{\end{eqnarray}}

\begin{document}
\title{Towards closing the window of primordial black holes as dark matter: \\ the case of large clustering}

\preprint{DESY 18-132}

\newcommand{\AddrCERN}{%
Theoretical Physics Department, CERN, Geneva, Switzerland
}
\newcommand{\AddrOslo}{%
Department of Physics, University of Oslo, Box 1048, N-0371 Oslo, Norway}
\newcommand{\AddrDESY}{%
Deutsches Elektronen-Synchrotron DESY,   Notkestra\ss e 85, D-22607 Hamburg, Germany}

 \author{Torsten Bringmann}
 \email{torsten.bringmann@fys.uio.no}
 \affiliation{\AddrOslo}
 
 \author{Paul Frederik Depta}
 \email{frederik.depta@desy.de}
 \affiliation{\AddrDESY}

\author{Valerie Domcke}
 \email{valerie.domcke@desy.de}
 \affiliation{\AddrDESY}

\author{Kai Schmidt-Hoberg}
 \email{kai.schmidt-hoberg@desy.de}
 \affiliation{\AddrDESY}

\begin{abstract}
The idea of dark matter in the form of primordial black holes has seen a recent revival triggered by the LIGO 
detection of gravitational waves from binary black hole mergers.
In this context, it has been argued that a large initial clustering of primordial black holes can help alleviate 
the strong constraints on this scenario. In this work, we show that on the contrary, 
with large initial clustering the problem is exacerbated and constraints on primordial black hole dark matter 
become overwhelmingly strong. 
\end{abstract}

\maketitle

\section{Introduction}%
Soon after realising that black holes (BHs) could form in the early radiation-dominated
universe \cite{1967SvA....10..602Z,Hawking:1971ei,Carr:1974nx} from the gravitational collapse 
of large 
density fluctuations, it was pointed out that such objects may even contribute 
appreciably to the total matter 
density \cite{1975Natur.253..251C}. An obvious question is therefore
 whether these primordial black holes (PBHs) could explain {\it all} of the 
cosmologically observed dark matter (DM), see Refs.~\cite{Carr:2016drx,Sasaki:2018dmp} for recent reviews. 
This idea has seen greatly renewed interest \cite{Bird:2016dcv,Clesse:2016vqa,Sasaki:2016jop,Carr:2017jsz,Wang:2016ana} after the discovery of 
binary mergers by the Advanced Laser Interferometer Gravitational-Wave 
Observatory (LIGO) \cite{Abbott:2016blz,Abbott:2016nmj,TheLIGOScientific:2016pea,Abbott:2017vtc},
proving the existence of $\mathcal{O}(10M_\odot)$ BHs with so far unclear origin.


Constraints on the allowed DM fraction $f_\mathrm{PBH}$ of PBHs derive from a large number 
of observations and have been explored for a vast range of mass scales, see 
Refs.~\cite{Carr:2016drx,Carr:2017jsz,Sasaki:2018dmp} for an overview.
While there seems to be a broad consensus that  $f_\mathrm{PBH}\sim1$ is essentially excluded for black hole masses $m_\text{PBH} \gtrsim 10^{-10}~M_\odot$ 
when assuming a homogeneously distributed population of PBHs with a  single mass, this general 
picture changes when either of these conditions is not met. Intriguingly, this also opens
the window of PBH masses consistent with the LIGO observations by circumventing the 
stringent constraints from microlensing and from the cosmic microwave background 
(CMB)~\cite{Garcia-Bellido:2017xvr} (see however~\cite{Zumalacarregui:2017qqd,Garcia-Bellido:2017imq}).

Clustered PBH distributions 
have been argued to arise generically in Refs.~\cite{Chisholm:2005vm,Chisholm:2011kn}, possibly explaining the 
existence of super-massive BHs~\cite{Dokuchaev:2004kr,Dokuchaev:2007mf}. Only recently 
it was realized that significant clustering is in fact {\it not} expected for Gaussian primordial 
fluctuations~\cite{Ali-Haimoud:2018dau,Desjacques:2018wuu}, as predicted by vanilla models of cosmic inflation. 
However, highly clustered PBH distributions could still plausibly arise, e.g., in the presence of sizeable primordial 
non-Gaussianities~\cite{Young:2014oea} or from the collapse of domain walls~\cite{Belotsky:2018wph}. 
In this work we choose to be agnostic about the possible origin of large PBH clustering. 
Instead we demonstrate that in such a situation $f_\text{PBH} \sim 1$ is in fact still excluded over a 
wide mass-range, thereby closing this possibly
last loop-hole for all DM consisting of PBHs with masses larger than $10^{-10} M_\odot$.


The initial clustering of PBHs is indeed a key parameter to understanding the 
phenomenology of PBH DM, affecting merger rates~\cite{Raidal:2017mfl,Ballesteros:2018swv}, 
the subsequent structure formation~\cite{Carr:2018rid}, and the interpretation of observational 
bounds~\cite{Garcia-Bellido:2017xvr}.
Here we take a pragmatic and phenomenological approach by parametrising the clustering as a constant, 
free parameter on the scales of interest. We 
point out that, for the large PBH clustering discussed in the literature, the expected merger rates easily 
exceed one per binary and Hubble time.  We demonstrate that
multiple subsequent mergers severely constrain PBH DM as a possible explanation of the LIGO events 
because of {\it i)} the expected (as compared to observed) merger rate,    
{\it ii)} the impact of the additional radiation component in gravitational waves (GWs) on both CMB 
and large-scale structure observations, and {\it iii)} a present-day stochastic GW background (SGWB) 
exceeding the sensitivities of current ground- (or future space-) based observatories. 

This article is organised as follows. We start by describing the GW spectrum and energy density
from cosmological PBH mergers, before discussing how the merger rate critically depends
on the initial PBH clustering. We then introduce a cascading merger scenario to capture the effects of large
clustering, and hence high merger rates. We derive the resulting contributions to the stochastic GW background and 
the relativistic energy density in GWs, using the cosmological parameters from Ref.~\cite{Aghanim:2018eyx} 
whenever relevant. Along with the actual event rate observed by LIGO, we use this to 
place constraints on $f_\mathrm{PBH}$. We discuss the influence of a deviation from the assumptions used in calculating the constraints and show that this does not qualitatively change our results.

\section{Gravitational waves from merging black holes}

Coalescing binary BHs emit GWs with a characteristic spectrum $\diff E_\mathrm{GW}/\diff \nu$, with
a total energy that makes up a significant fraction of the rest 
mass \cite{Maggiore:2018sht}. For BHs with identical masses and negligible spins, e.g., one
expects $E_\mathrm{GW} / M_\mathrm{2BH} \simeq 5\%$, where $M_\mathrm{2BH}$ is the initial mass of the
system. For the ten events observed so far by LIGO, this number ranges between $3.3\%$ and $5.4\%$ \cite{LIGOScientific:2018mvr}.

In the following we will study the cumulative effect of all mergers throughout the cosmological evolution. 
The resulting present energy density {\it per logarithmic 
frequency interval} is conventionally expressed in units of the critical density, $\rho_c=3H_0^2/(8\pi G)$, 
and computed as \cite{Phinney:2001di,TheLIGOScientific:2016wyq}
\be
\Omega_\text{GW} (\nu) = \frac{1}{\rho_c}\int_0^\infty \!\!\!\!\diff \tilde{z} \int  \!\diff R(\tilde{z}) \frac{\nu}{(1+\tilde{z}) H(\tilde{z})} \frac{\diff E_\text{GW}}{\diff \nu_s}\,.
\label{eq:bg}
\ee
Here, the merger rate is denoted as $R(z)$, where $z$ is the cosmological redshift, 
$H$ is the Hubble rate, and
 the observed frequency $\nu$ corresponds to an emission frequency of ${\nu_s = \nu (1+z)}$.
The {\it total} energy density in gravitational waves at any given redshift $z\equiv a-1$ is therefore  
\be
\frac{\rho_\text{GW} (z)}{(1+z)^4}= \int_z^\infty  \!\!\!\!\diff \tilde{z} \int\! \diff \nu \!\int\! \diff R(\tilde{z}) \frac{\diff E_\mathrm{GW}/\diff \nu}{(1+\tilde{z})^2 H(\tilde{z})} \,. \label{eq:rhoGW}
\ee
Covariant conservation of energy implies that the mass density in PBHs must correspondingly decrease as 
$a^{-3}\diff(a^3 \rho_\mathrm{PBH})\!=\!-a^{-4}\diff(a^4 \rho_\mathrm{GW})$ \cite{Bringmann:2018jpr}, which
gives
\begin{align}
&\frac{\rho_\text{PBH}(z)}{(1+z)^3} =C - \int_z^\infty  \!\!\!\!\diff \tilde{z} \int\! \diff \nu \!\int\! \diff R(\tilde{z}) 
\frac{\diff E_\mathrm{GW}/\diff \nu}{(1+\tilde{z}) H(\tilde{z})}\,.\label{eq:rhoPBH}
\end{align}
We fix the integration constant $C$ 
such that $f_\mathrm{PBH}\equiv \left(\rho_\mathrm{PBH}/\rho_\mathrm{DM}\right)_{z_\mathrm{CMB}}$ 
is the PBH fraction at \mbox{$z={z_\mathrm{CMB}}\simeq1100$}.
For the spectrum $\diff E_\mathrm{GW}/\diff \nu$ we use commonly 
adopted fitting formulae~\cite{Cutler:1993vq,Zhu:2011bd,Chernoff:1993th,Ajith:2007kx}.

\section{Merger rates and clustering}%
In the early universe, PBH binary formation starts once the Newtonian force between 
two initial PBHs overcomes the Hubble flow, with a nearby third PBH providing the angular momentum 
necessary to prevent a head-on collision \cite{Nakamura:1997sm,Ioka:1998nz,Sasaki:2016jop}. Later, 
peculiar velocities may be too large for this to happen; instead, binary formation can be triggered
by the energy loss in GWs during close encounters of two PBHs \cite{1989ApJ...343..725Q,Mouri:2002mc}.
For the parameter combinations of interest to us, though, the rate associated with the first formation mechanism by far 
exceeds that for the second, even for binaries merging only today~\cite{Raidal:2017mfl}. Once formed, these 
binary systems survive until they merge, 
largely unaffected by the evolution of the surrounding Universe~\cite{Ali-Haimoud:2017rtz}.

A crucial input for calculating those merger rates is the initial clustering of PBHs. Phenomenologically,
this can be described in terms of an idealised two-point correlation function $\xi_\mathrm{PBH}(r)$ that is 
constant at scales relevant for the formation of PBH binaries \cite{Raidal:2017mfl}: 
\be
  1+\xi_\mathrm{PBH}(r)\approx \delta_\mathrm{dc} = const.\,,
\ee
where $\delta_\mathrm{dc}$
 describes the local density contrast, evaluated at the time when the two BHs decouple
from the Hubble expansion. A perfectly homogeneous PBH distribution corresponds to 
$\delta_\mathrm{dc}=1$, while a highly clustered PBH distribution is described by $\delta_\mathrm{dc}\gg1$. 
Values of $\delta_{\text{dc}} \gtrsim 10^5$ are particularly interesting, as they are required to circumvent the tight constraints on PBH 
DM from microlensing and the CMB~\cite{Garcia-Bellido:2017xvr}. Furthermore, constraints arising 
from the conversion of PBH DM into gravitational radiation are alleviated if PBH mergers occur only at high 
redshift, which was demonstrated to happen for  $\delta_\text{dc} \gtrsim 10^4$ assuming 
only a single merger step  \cite{Raidal:2017mfl}.

We will see that a highly clustered initial PBH population not only leads to a 
more efficient formation
of binaries but also to a 
significantly enhanced merger rate, during the whole cosmological evolution until today. These findings  crucially extend
previous results in the literature (see e.g.\ \cite{Raidal:2017mfl,Ballesteros:2018swv}), which consider 
the impact of clustering only on a single merger step. 

\section{Cascading black hole merger events}%
We thus need to improve these scenarios 
by allowing for subsequent merger steps, i.e.\ binary mergers of systems of previously merged PBHs.
For simplicity, we model the PBHS distribution as an
{\it initially} monochromatic mass distribution peaked at $m_0$ and assume that 
the PBH masses in merger step $j$ are given by
\begin{align}
 m_j & = 2 m_{j - 1} - E_\text{GW}(m_{j-1}) 
  \sim 1.9^j \, m_0 \,. \label{eq:mass}
\end{align}
The impact of more realistic, extended initial mass functions is discussed in Sec.~\ref{sec:discussion}, where 
we demonstrate that our key results are not affected by this choice.
The average PBH number density $n_j = \rho_{\text{PBH}, \infty}/(2^j m_0)$  is locally enhanced by a 
factor of $1 + \xi(r) \approx \delta_{\text{dc},j}$,  where $\rho_{\text{PBH}, \infty}$ denotes the initial PBH density 
and we model the decrease of the PBH clustering as
$\delta_{\text{dc},j} \simeq 2^{-j}\, \delta_{\text{dc},0}$. The details of this decrease will in general
depend on the exact form of the two-point correlation function; as argued in Sec.~\ref{sec:discussion}, the specific choice above leads to rather conservative limits on the fraction of PBH DM.

The merger rate of the $j$th merger step is given by
\begin{align}
R_{j} (t) = &\int_0^{\tilde{x}} \diff x \int_{\tilde{x}}^\infty \diff y \,  \frac{\partial^2 n_{3,j} (x,y)}{\partial x \partial y} \nonumber \\
&\delta (t - \tau (x,y,m_{j}) - \max (t_{\mathrm{dc},j} (x), t_\mathrm{form}) )\,, \label{eq:R3jDef}
\end{align}
where $x$ and $y$ denote the comoving distances from a given PBH to the nearest and next-to-nearest PBH, 
respectively, and the number density of PBH triples $n_{3,j}(x,y)$ is given by~\cite{Raidal:2017mfl}
\begin{align}
\!\!\diff n_{3,j} (x,y) = \frac{n_j}{2} \mathrm{e}^{-\frac{4 \pi}{3} y^3 n_j \delta_{\mathrm{dc},j}} (4 \pi n_j \delta_{\mathrm{dc},j})^2 x^2 y^2 \diff x \diff y\,.\!\!
\label{eq:dn3}
\end{align}
The delta distribution in Eq.~\eqref{eq:R3jDef} ensures that the coalescence time $\tau$~\cite{Peters:1964zz},
\begin{align}
\tau (x,y,m_{j}) = \tilde{\tau}_j \left({x}/{\tilde{x}_j} \right)^{37} \left( {y}/{\tilde{x}_j} \right)^{-21}\,,
\end{align}
with
\begin{align}
\tilde{\tau}_j = \frac{3 a_\mathrm{eq}^4 \tilde{x}_j^4}{170 (G m_{j})^3}\,, \qquad 
\tilde{x}_j^3 = \frac{3}{4 \pi} \frac{2 m_{j}}{a_\text{eq}^3 \rho_\text{eq}}\,,
\end{align}
is measured from when the PBHs are both formed ($t_\text{form}$) and decoupled from the Hubble 
flow ($t_{\mathrm{dc},j}$). PBHs form almost immediately after the corresponding density perturbations enter 
the Hubble horizon, with a mass $m_0$ equalling the total energy within the horizon at that time, 
and the decoupling from the Hubble flow occurs when the gravitational attraction overcomes the Hubble expansion:
\begin{equation}
 t_\text{form} = G m_0 \,, \quad t_{\text{dc},j} = \left( \frac{16 \pi G}{3} \rho_\text{eq} \right)^{-1/2} \left( \frac{x}{\tilde x_j} \right)^6 \,.
\end{equation}
The subscript `eq' above refers to matter radiation equality. 
The merger rates $R_j$  in Eq.~\eqref{eq:R3jDef} are connected to the differential one employed in Eqs.~\eqref{eq:bg}-\eqref{eq:rhoPBH} via
\begin{equation}
\diff R (\tilde{z}) = \sum_j R_j (t(\tilde{z}))\, \delta (m - m_j) \diff m\,.
\end{equation}

To recap, we consider a scenario of subsequent equal-mass mergers with a corresponding shift in the 
mass distribution and local density contrast in each merger step. Let us stress that even though there are
characteristic time-scales implied by the merger rates, we allow PBHs of given mass $m_j$ to merge 
at any time between the decoupling for the $j$th step (as long as $t_{\text{dc},j}>t_\text{form}$) and today. 
For rare very early mergers this may lead to situations where coalescence in our model begins when, in reality, 
instead of the eventually merging two PBHs a preceding set of smaller PBHs is present; in this
case, we slightly underestimate the actual amount of emitted GWs.
We further introduce an approximate scenario to better visualise the individual merger steps,
adopting rates $R_j$ that are
zero before $\max (t_{\text{dc},j}, t_\mathrm{form})$, constant until 
the average coalescence time has passed (with average values for $x$ and $y$), and zero afterwards.

 \begin{figure}[t]
 \includegraphics[width=\columnwidth]{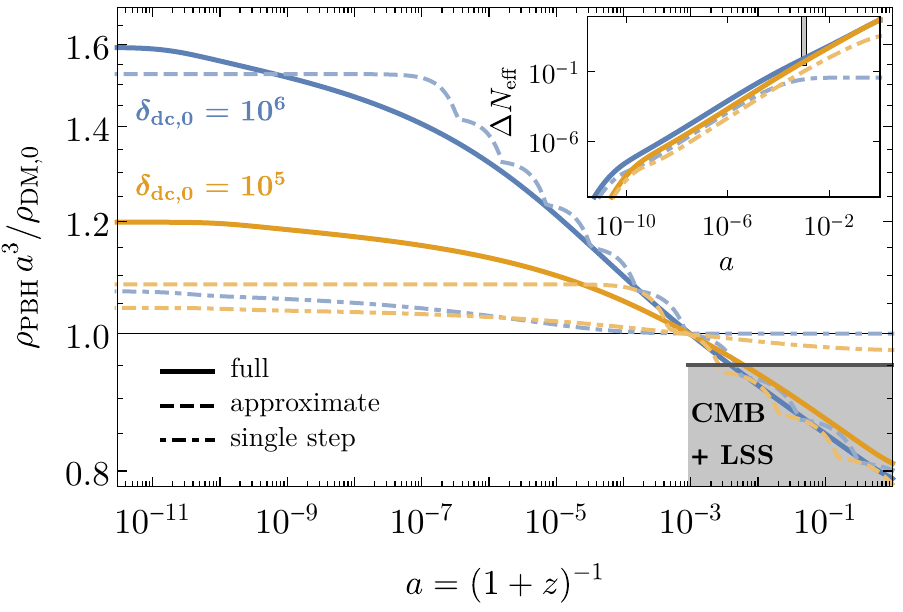}
 \caption{Conversion of PBH energy density (main panel) into gravitational wave radiation (in terms of 
 a corresponding number $\Delta N_\mathrm{eff}$ of additional neutrinos, top right inlet) for $m_{0} = 1 M_\odot$, 
 $f_\mathrm{PBH}=1$, and different theoretical treatments. In grey, we indicate regions excluded by cosmology \cite{Bringmann:2018jpr}.
 \label{fig:rho_cascade}}
 \end{figure}

In Fig.~\ref{fig:rho_cascade} we show the decrease of the PBH energy density in our merger scenario (solid lines)
as well as the corresponding increase in gravitational wave radiation (top right inlet). 
Overall, the agreement between the full rates, as computed from Eq.~\eqref{eq:R3jDef}, 
and the simple approximate scenario mentioned above (dashed lines) is fairly good. 
In particular we note the sizeable equidistant spacing of the merger steps, 
justifying the assumption of a hierarchical merging scenario.
The dash-dotted curves represent the common assumption of a single merger step. For larger values of 
$\delta_{\text{dc},0}$, this merger step occurs earlier \cite{Raidal:2017mfl}, leading to the tempting conclusion 
that bounds on the GW production can be evaded since the produced radiation is highly red-shifted. As we discuss
below, this conclusion clearly no longer holds once multiple merger steps are taken into account.

\section{Cosmological bounds}%
The conversion of PBH DM into GW radiation modifies the standard cosmological evolution and is 
constrained by CMB and large scale structure (LSS) observations~\cite{Bringmann:2018jpr}. 
These modifications can be roughly split into {\it i)} an upper bound on the effective number of neutrino species at the time 
of the CMB, $\Delta N_\text{eff}(z_\text{CMB}) \lesssim 0.3$~\cite{Aghanim:2018eyx}, indicated by the grey area
in the inlet of Fig.~\ref{fig:rho_cascade}, and {\it ii)} the amount of 
DM converted into invisible (`dark') radiation at later times. From Fig.~6 in Ref.~\cite{Bringmann:2018jpr} one can deduce that 
not more than $\sim 5\%$ of DM can be converted into dark radiation after the CMB epoch, irrespective of the 
precise time-dependence of this conversion (and consistent with the
$4.2 \%$ found for the case of decaying DM~\cite{Poulin:2016nat}). For simplicity, and because other constraints 
turn out to be stronger, we conservatively adopt this bound of 5\% in our analysis (indicated 
by the grey region 
in the bottom right of Fig.~\ref{fig:rho_cascade}). 

\begin{figure}[t]
\includegraphics[width=\columnwidth]{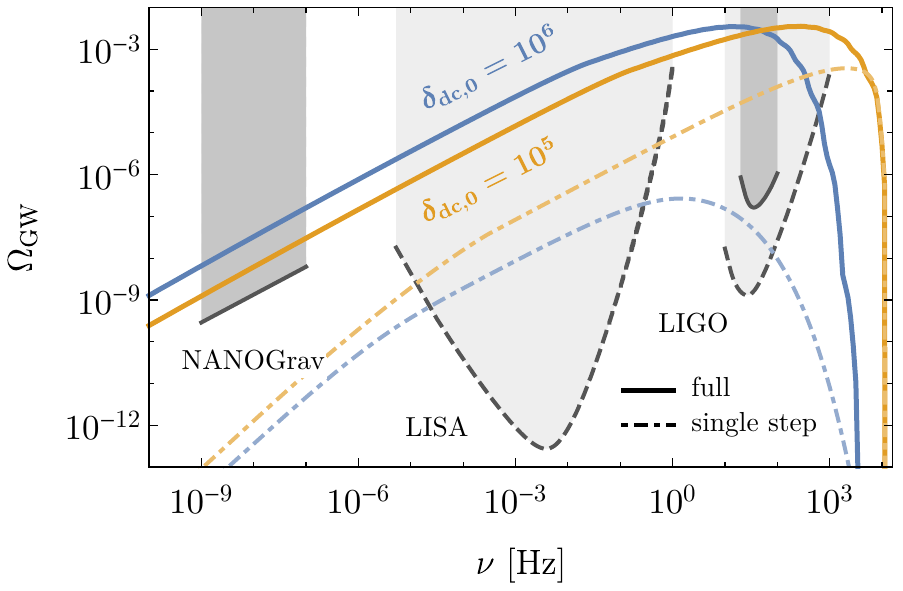}
\caption{GW density parameter per logarithmic frequency interval for $m_{0} = 1~M_\odot$, $f_\mathrm{PBH}=1$, and 
different theoretical treatments. In grey, we indicate present (solid lower lines) and projected (dashed lower lines) 
constraints from NANOGrav~\cite{Arzoumanian:2018saf}, LISA~\cite{Audley:2017drz}, and 
LIGO~\cite{TheLIGOScientific:2016dpb} (present O1 and projected design constraint).
\label{fig:GW_BG}}
\end{figure}

\section{The stochastic gravitational wave background}%
With the cumulative merger rate described above, it is also straightforward to compute the resulting SGWB as given in 
Eq.~(\ref{eq:bg}). In Fig.~\ref{fig:GW_BG}, we illustrate the predictions
for an initial PBH mass of $1~M_\odot$ and different initial clusterings $\delta_\mathrm{dc,0}=10^5$ 
and $\delta_\mathrm{dc,0}=10^6$.
The GW spectrum is dominated by late time mergers ($z \lesssim$ 10), since earlier GW emission is highly diluted by 
cosmic expansion. Larger clustering implies that most of these late mergers are associated with heavier BHs, 
which emit GWs with lower frequencies.
For the parameter example of Fig.~\ref{fig:GW_BG} we find that  the mergers occurring at $z = 0$ have typically 
undergone 4 (9) previous mergers for $\delta_\text{dc,0} = 10^5 (10^6)$, resulting in a PBH mass today of around 13 (300) $M_\odot$. 
Since the frequency
of the emitted GWs roughly scales as $\nu \sim 1/M_{2\text{BH}}$, cf.~\cite{Raidal:2017mfl, Maggiore:1900zz}, 
this explains the shift between the two solid lines in Fig.~\ref{fig:GW_BG}.
Moreover, the merger cascade described above leads to a mild broadening of the 
high frequency peak. At low frequencies, the $\Omega_\text{GW} \sim \nu^{2/3}$ scaling indicates the early inspiral 
phase of the BH binaries \cite{Zhu:2011bd}. 

For comparison, the dash-dotted curves show the predictions for a single merger step, where  
the main effect of large clustering is to shift the merging time to high redshift, strongly suppressing the GW 
spectrum.
However, as Fig.~\ref{fig:GW_BG} demonstrates, later mergers completely change the picture, leading to a large contribution to the SGWB.
The grey contours, finally, indicate the power-law integrated sensitivity curves of LIGO \cite{TheLIGOScientific:2016dpb} 
and the pulsar timing array NANOGrav~\cite{Arzoumanian:2018saf}, as well as the planned space-based 
LISA~\cite{Audley:2017drz} observatory. 

 \setlength{\tabcolsep}{5pt}
 \begin{table}[t]
  \begin{tabular}{c | cccccccc}
   $m/M_\odot$ & 0.2 & 1 & 10 & 20 & 40 & 100 & 200 & 300 \\
   $R \, \text{Gpc}^3  \, \text{yr}$ & $10^6$ & $1.9 \cdot 10^4$ & 330 & 77 & 15 & 2 & 5 & 20
  \end{tabular}
 \caption{$90\%$ CL upper limits on merger rates in the late universe, taken from Refs.~\cite{Abbott:2016nhf,Abbott:2016drs,Abbott:2017iws,Abbott:2018oah}.}
 \label{tab:rates}
 \end{table}

 \section{Observed merger rate}
 The 
 LIGO/VIRGO observations strongly constrain the merger rate of PBHs with masses
 between 0.2 and 300 $M_\odot$~\cite{Abbott:2016nhf,Abbott:2016drs,Abbott:2017iws,Abbott:2018oah}. 
 Interpolating linearly between the limiting rates stated in Tab.~\ref{tab:rates}, 
 and comparing this to the calculated $R_j (z=0)$, allows us to derive an upper bound on $f_\text{PBH}$.
 Starting with an initially monochromatic mass function, we would not expect to reproduce the BH mass distribution observed by LIGO. However, requiring to reproduce the total observed merger rate ($12-213~\text{Gpc}^{-3} \text{yr}^{-1}$~\cite{Abbott:2017vtc}) with PBH mergers in the sensitivity band of LIGO ($7 - 50 M_\odot$), we obtain a (very conservative) range in $m_0$ compatible with the total merger rate observed by LIGO.

\begin{figure*}[t]
\includegraphics[width=\columnwidth]{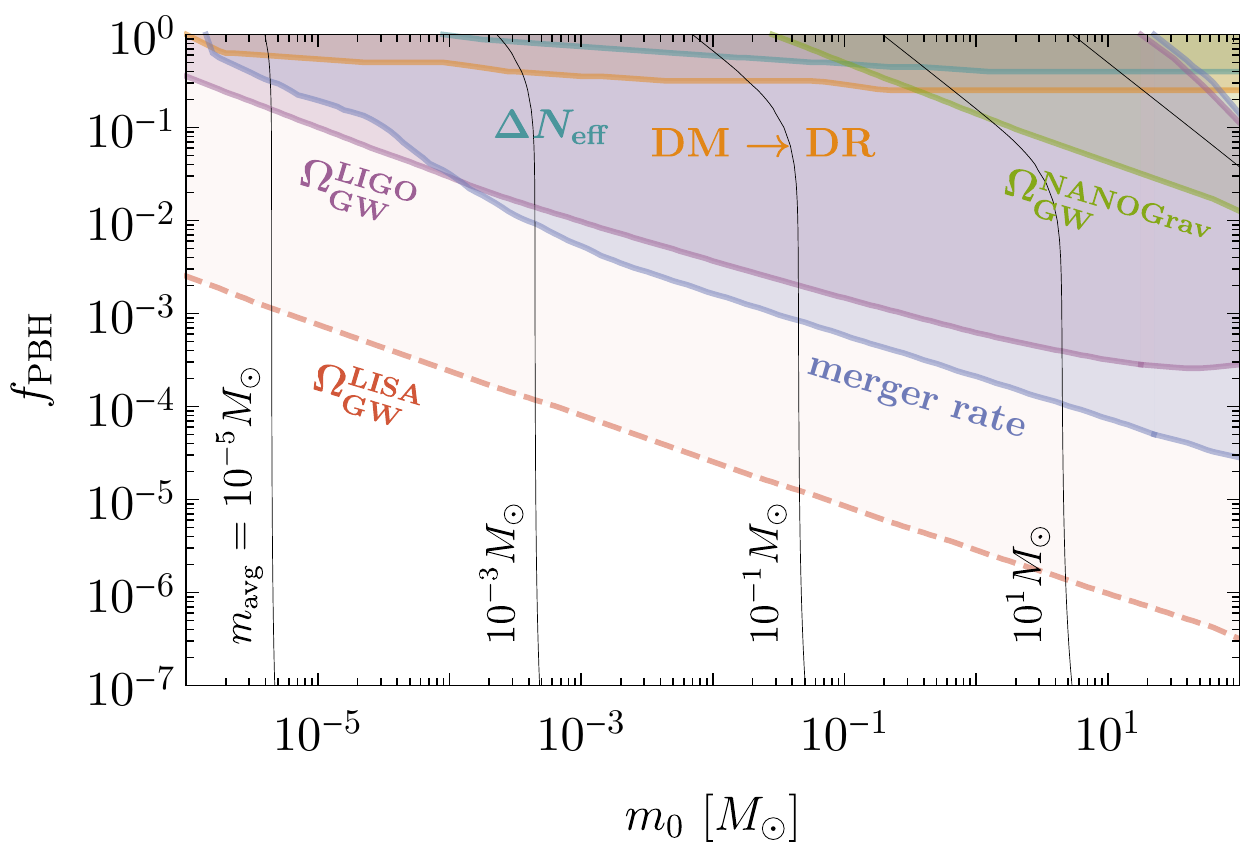}~~~
\includegraphics[width=\columnwidth]{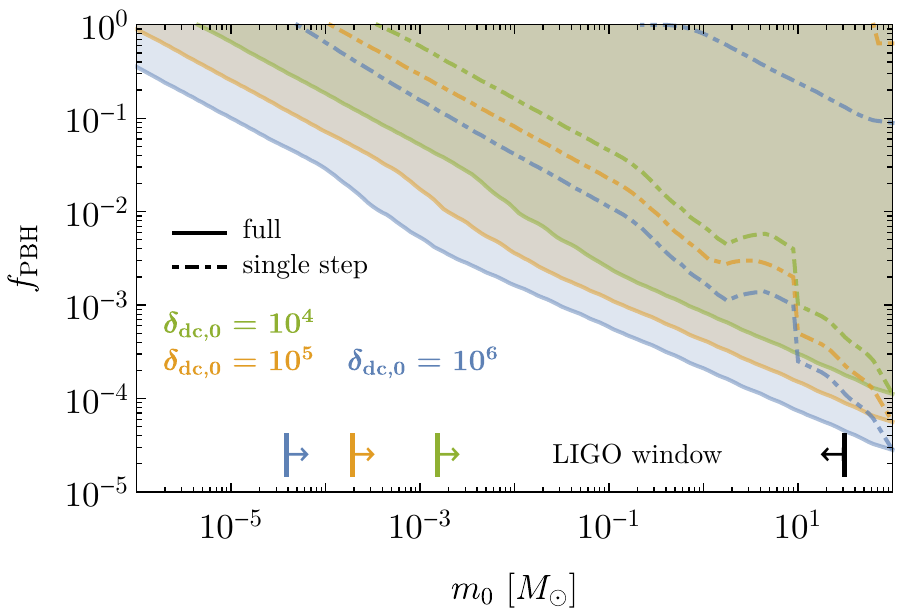}
\caption{%
{\it Left.} Constraints on the allowed fraction $f_\text{PBH}$ of PBH DM  as a function of the initial PBH mass $m_0$ for 
large clustering ($\delta_\text{dc,0} = 10^6$) in the merger cascade scenario. 
The thinner black lines indicate contours of the rate-averaged PBH mass $m_\mathrm{avg} = (\sum_j R_j m_j)/\sum_j R_j$ at $z = 0$.
{\it Right.} Combined constraints on $f_\text{PBH}$ for different clustering parameters $\delta_{\text{dc},0}$.
\label{fig:constraints}}
\end{figure*}
\section{Results}%
In the left panel of Fig.~\ref{fig:constraints}, we summarise the resulting constraints on the allowed fraction of DM in PBHs 
for large initial clustering, $\delta_\text{dc,0} = 10^6$, as a function of the initial PBH mass $m_0$ (the shaded regions
are excluded). For reference, we also indicate contour lines with the present, rate-averaged PBH mass $m_{\rm avg}$.
We depict as blue-green and orange curves, respectively, the cosmological constraints~\cite{Bringmann:2018jpr}
indicated as grey shaded areas in Fig.~\ref{fig:rho_cascade}. The blue solid line shows the merger rate constraint;
we note that it extends to {\it average} PBH masses well below the LIGO/VIRGO limit because a small fraction of PBHs
will still satisfy $m_{\rm PBH}>0.2\,M_\odot$ after many merger steps.
The remaining lines, finally, correspond to the SGWB constraints from NANOGrav (green) and LIGO (purple) 
indicated in Fig.~\ref{fig:GW_BG}.
 The upcoming space-based LISA experiment (red, dashed) may severely tighten these constraints.

In the right panel of Fig.~\ref{fig:constraints}, we show our combined results on $f_\text{PBH}$, illustrating that larger values of 
the clustering parameter $\delta_{\text{dc},0}$ in fact lead to tighter constraints. 
For comparison, the dash-dotted lines indicate the much weaker constraints obtained when taking into account only a single merger step. The arrows indicate the range for $m_0$,
where for a suitable $f_\text{PBH}$ the total present merger rate is consistent with all observed LIGO events being caused by PBH mergers.

\section{Discussion}%
\label{sec:discussion}
We made a number of simplifying assumptions when modelling the PBH merger history, 
which can impact the details of the limits summarized in Fig.~\ref{fig:constraints}.
As we argue below, however, our main result that a large clustering leads to a {\it tightening}
of existing limits is robust. This implies, in particular, that the BHs observed by LIGO cannot
be part of a PBH population comprising all of the DM.


\subsection{Extended mass distributions}

In the above calculation, we focused for simplicity on an initially monochromatic PBH mass distribution. Realistic models of PBH formation however predict extended mass functions. For example, if the scalar perturbations which collapse into PBHs are formed during single-field inflation, then the characteristic time-scale governing the width of any peaked feature in the scalar power spectrum is the Hubble time. Consequently, the resulting PBH mass distribution is expected to be rather broad, since the range of enhanced length scales is exponentially sensitive to this time-scale~\cite{Carr:2009jm}   
(though notable exceptions exist \cite{Starobinsky:1992ts,Blais:2002gw,Blais:2002nd, Orlofsky:2016vbd}). More generally, even if the spectrum of enhanced scalar perturbations is essentially monochromatic, the process of collapsing the initial perturbations into PBHs obeys a critical scaling relation, implying a finite width for the resulting PBH mass spectrum~\cite{Carr:2016drx}. Here we discuss how the results in the main text are affected by relaxing the assumptions of an initially monochromatic PBH mass distribution.

To model an extended PBH mass function, we consider the lognormal mass distribution
\begin{align}
\psi_\mathrm{log} (m) = \frac{f_\mathrm{PBH,\infty}}{\sqrt{2 \pi} \sigma m} \exp \left( \frac{\log^2 (m/m_0)}{2 \sigma^2} \right)\,, 
\label{eq:lognormal}
\end{align}
where $\sigma$ encodes the width of the distribution centered around the reference scale $m_0$. This distribution 
is normalized so that $f_\mathrm{PBH,\infty} = \int_0^\infty\! \diff m \, \psi (m)$. Note that for computational simplicity, we fix the PBH fraction of DM at $z \rightarrow \infty$ in this Section as indicated by the subscript $\infty$. In 
our scenario of multiple mergers, the strongest constraints on the PBH abundance arise from the bounds on the 
SGWB and from the observed merger rate in LIGO, see Fig.~\ref{fig:constraints}. Consequently we will focus on 
the impact of the distribution~\eqref{eq:lognormal} on these two quantities, noting that in particular significant 
changes in the (hierarchy of) the merger rates at the individual steps could potentially entail significant changes 
in our computations.

\begin{figure}[t]
	\vspace{-0.3cm}
	\includegraphics[width=\columnwidth]{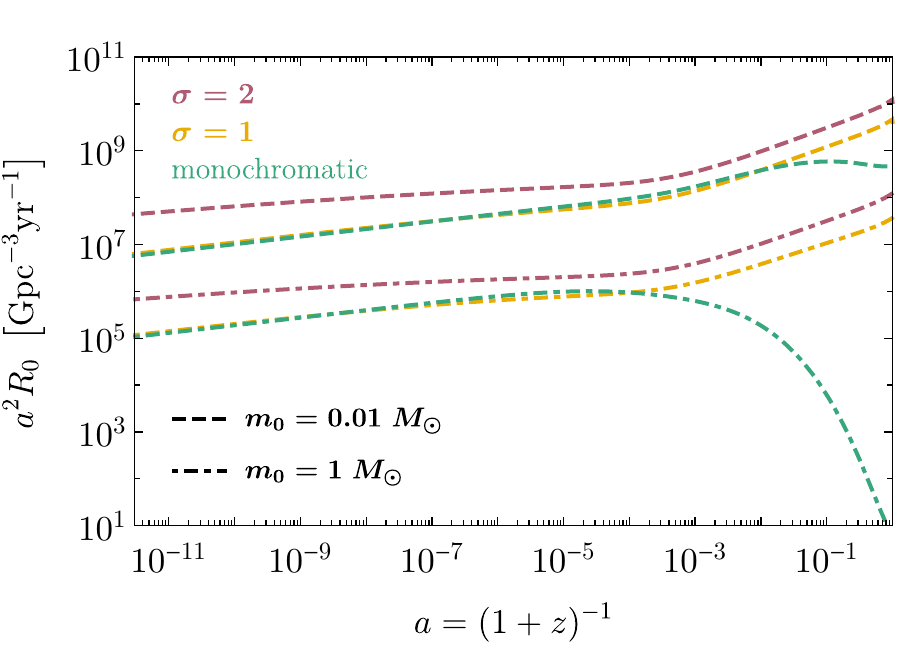}
	\caption{%
		Evolution of mass-integrated merger rate for lognormal mass distributions with $\sigma = 1$ and $2$ and monochromatic mass distributions considering only one step with $f_\mathrm{PBH} = 1$, $\delta_{\mathrm{dc},0} = 10^6$. Note the additional factor of $a^2$ on the $y$-axis.
		\label{fig:rateExtMassDistr}}
\end{figure}

Fig.~\ref{fig:rateExtMassDistr} shows the time evolution of the merger rate (integrated over all masses) for different mass distributions for a {\it single} merger step, obtained by re-instating an extended mass function in Eq.~\eqref{eq:R3jDef}. 
Here, the coalescence time depends on the masses of the PBHs in the binary ($m^{(1)}$ and $m^{(2)}$) and the third PBH ($m^{(3)}$) providing angular momentum, and
the number density of PBH triples is now given by
\begin{align}
\diff n_{3,0}^{\mathrm{log/mono}} (x,y) = &\frac{\rho_\mathrm{PBH,\infty}}{2} \mathrm{e}^{- N(y)} (4 \pi \rho_\mathrm{PBH,\infty} \delta_{\mathrm{dc},0})^2 \nonumber \\
&x^2 y^2 \diff x \diff y \prod_{i=1}^{3} \frac{\psi_\mathrm{log/mono}}{m^{(i)}} \diff m^{(i)}\,,
\label{eq:dn3ExtMassDist}
\end{align}
where
\begin{align}
\rho_\mathrm{PBH,\infty} &= \lim_{z \rightarrow \infty} (1+z)^{-3} \rho_\mathrm{PBH} (z)\,, \\
N (y) &= \int_0^y\! \diff \tilde{y} \int_0^m\! \diff \tilde{m} \nonumber \\
&\quad \; 4 \pi \tilde{y}^2 \delta_{\mathrm{dc},0} \rho_\mathrm{PBH,\infty} \frac{\psi_\mathrm{log/mono} (\tilde{m})}{\tilde{m}}\,, \\
\psi_\mathrm{mono} (m) &= f_\mathrm{PBH,\infty} \delta (m - m_0)\,.
\end{align}
For more details see~\cite{Raidal:2017mfl}. We note that the time evolution of this merger rate does not change 
significantly when going from a monochromatic to a lognormal mass distribution. This situation only changes at 
very late times, when the merger rate of this single merger step has however anyway dropped to a small value (note the 
additional factor of $a^2$ in Fig.~\ref{fig:rateExtMassDistr}).
For the density contrast employed in Fig.~\ref{fig:rateExtMassDistr}, $\delta_{\mathrm{dc},0} = 10^6$, subsequent steps start to dominate the total merger rate for $a \gtrsim 5 \cdot 10^{-5}$ and $a \gtrsim 10^{-6}$ for $m_0 = 0.01~M_\odot$ and $m_0 = 1~M_\odot$, respectively.
In conclusion, both the hierarchy among the 
merger times of the individual merger steps as well as the total amount of mergers are not significantly altered 
when assuming a rather narrow lognormal mass distribution. For broader distributions the rate is even enhanced.

The SGWB for the lognormal distribution~\eqref{eq:lognormal} was studied in~\cite{Raidal:2017mfl} for a single 
merger step, finding (for $\sigma = 1$) a similar GW spectrum as in the monochromatic case. Since the GW 
radiation from early PBH mergers experiences a significant redshift, the SGWB in the multiple merger scenario 
is dominated by the very last merger. Consequently, since we found above that the overall merger history is not 
significantly altered in the case of a lognormal mass distribution, we conclude that also the resulting SGWB is only moderately altered.

In summary, the strongest constraints depicted in Fig.~\ref{fig:constraints} do not appear to be very sensitive to the assumption of a monochromatic initial PBH mass function. Depending on the mass function in question, they may vary by a factor of a few or even an order of magnitude, but the option of obtaining a significant PBH dark matter fraction in the LIGO window of $\sim 10~M_\odot$ remains firmly excluded. For a given peaked mass distribution one may conservatively estimate the constraints on highly clustered PBH dark matter by considering a narrow window around the central mass value, and applying the analysis of the main text to this window only.

\subsection{Decrease of local density contrast}

An important parameter in the modelling of the multiple merger scenario is the decrease in the local density contrast 
$\delta_\text{dc}$, which depends on the details of the two-point correlation function. In a 
large PBH `cloud' of constant density, the (number) density contrast stays approximately constant through many merger 
steps. On the contrary, for PBH `clouds' in which the typical distance from one PBH binary to the next is enhanced
compared to the initial size of the binary, the density contrast drops faster. In the above calculation, we 
considered  $\delta_\mathrm{dc,j} = 0.5^j \delta_\mathrm{dc,0}$, which we consider a conservative 
approach for highly clustered PBH distributions. 

\begin{figure}[t]
	\includegraphics[width=\columnwidth]{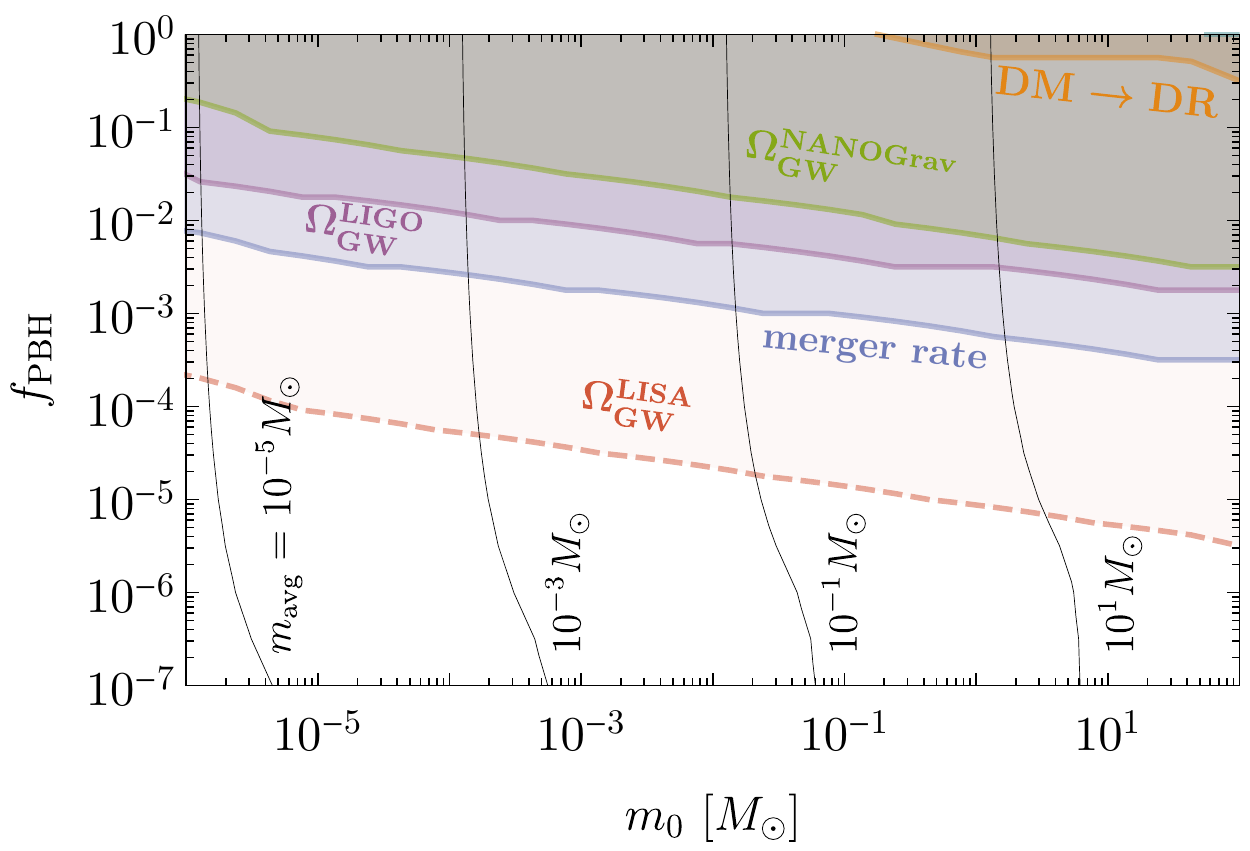}
	\caption{%
		Constraints for smaller initial clustering, $\delta_\mathrm{dc,0} = 10^3$ and $\delta_\mathrm{dc,j} = 0.9^j \delta_{\mathrm{dc},0}$. See~Fig.~\ref{fig:constraints} for details.
		\label{fig:constraintsDiffDecreases}}
\end{figure}

Here we also discuss the case of a locally approximately homogeneous density contrast.
To model this, we consider 
$\delta_\mathrm{dc,j} = 0.9^j \delta_\mathrm{dc,0}$. The resulting constraints, depicted in Fig.~\ref{fig:constraintsDiffDecreases} 
for $\delta_\text{dc,0} = 10^3$, are clearly significantly tighter than those presented in Fig.~\ref{fig:constraints}. Note 
that in this case the bounds can be extrapolated down to very low masses. Taken at face value, this 
even allows to constrain masses as low as  $m_0 = 10^{-16} M_\odot$, but this extrapolation implicitly 
assumes very large overdensities of $\mathcal{O} (10^{15})$ PBHs, which seems hardly realistic. We 
also note that such a slow decrease in the density contrast weakens the hierarchy between the merger 
times of the individual merger steps, making our analysis less robust.

We conclude this section by noting that for this slower decrease in the density contrast corresponding to rather homogeneous PBH `clouds', the formation of $N$-body systems may lead to a more complicated merger history not covered by the treatment in this work~\cite{Raidal:2018bbj}. 


\subsection{Gravitational wave recoil}

Due to asymmetries in the emitted gravitational radiation, the remnant of merging binary BHs can gain recoil 
velocities of $\mathcal{O} (100 - 1000~\text{km/s})$~\cite{Koppitz:2007ev,Rezzolla:2007xa,2011PhRvL.107w1102L}. These velocities are 
zero in the equal-mass zero-spin case, but differ from zero if either of these conditions are not met. Specifically, 
a maximal velocity of $\sim 175~\text{km/s}$ can be obtained in the zero-spin case for a BH mass ratio of $0.36$. 
For equal masses and non-zero spins with special alignment the maximum recoil velocity can be as large as 
$\sim 5000~\text{km/s}$~\cite{2011PhRvL.107w1102L}. As special spin alignment is unlikely in the context of this work, we expect recoil velocities of $\mathcal{O} (100~\text{km/s})$. In our treatment of multiple merger steps we 
have implicitly assumed that recoils do not affect the evolution of the PBHs. Here we 
show that this approximation is justified, by comparing these recoil velocities to the relevant escape velocities.

The escape velocity from a PBH cluster will depend on its global structure and hence on the two-point 
correlation function. As we take the local density contrast to decrease as 
$\delta_{\mathrm{dc},j} \simeq 2^{-j} \delta_{\mathrm{dc},0}$, we expect initial clusters to contain 
$\sim \delta_\mathrm{dc,0}$ PBHs with comoving radius $x_\mathrm{clus} = ( 3/4 \pi n_0 )^{1/3}$, 
such that $4 \pi x_\mathrm{clus}^3 n_0 \delta_\mathrm{dc,0}/3 = \delta_\mathrm{dc,0}$.
These clusters decouple from the Hubble flow during radiation domination when the local energy 
density in PBHs $\rho_\mathrm{PBH} \delta_{\mathrm{dc},0}$ is equal to the background density. 
This gives a decoupling scale factor and an escape velocity from the cluster of
\begin{align}
a_\mathrm{dc,clus} &= a_\mathrm{eq}^4 \frac{\rho_\mathrm{eq}}{2 m_0 n_0 \delta_\mathrm{dc,0}}\,, \\
v_\mathrm{esc,clus} &\simeq \sqrt{\frac{2 G \delta_\mathrm{dc,0} m_0}{x_\mathrm{clus} a_\mathrm{dc,clus}}} \nonumber \\
&= 3600~\frac{\text{km}}{\text{s}} f_\mathrm{PBH}^{2/3} \left( \frac{m_0}{M_\odot} \right)^{1/3} \frac{\delta_\mathrm{dc,0}}{10^4}\,. \label{eq:vEscClus}
\end{align}
This implies that for most regions of parameter space recoils should not affect our results. 

This conclusion may seemingly be challenged for small initial masses and small values of $f_\mathrm{PBH}$.
In this case, however, our constraints are driven by rare events of many merger steps, implying that initially the 
PBHs were exceptionally close. Specifically note that any mergers we take into account in our limits
must have a coalescence time $\tau$ smaller than the age of the universe $t_0$, i.e.\
\begin{align}
\tau = \tilde{\tau}_j \left({x}/{\tilde{x}_j} \right)^{37} \left( {y}/{\tilde{x}_j} \right)^{-21} \leq t_0\,.
\end{align}
Conservatively estimating $y \simeq 2^{1/3} x$ (smaller values give larger escape velocities and larger 
values are exponentially suppressed by the exponential term in Eq.~\eqref{eq:dn3}), the escape 
velocity from a binary BH in merger step $j$ can be written as
\begin{align}
v_\mathrm{esc,bin} = \sqrt{\frac{2 G m_j}{x a_{\mathrm{dc},j}}} \gtrsim 190~\frac{\text{km}}{\text{s}} \left( \frac{m_j}{M_\odot} \right)^{1/8}\,, \label{vEscBin}
\end{align}
where $a_{\mathrm{dc},j}$ is the scale factor at which the binary system decouples from the 
Hubble flow. We see that PBHs with $m_j \gtrsim 0.1 M_\odot$, which are relevant for the 
merger rate and the stochastic GW background from LIGO, typically do not pick up a 
sufficient recoil in their previous merger to escape the gravitational attraction of even their nearest neighbour.
Note however that even sub-critical recoils as well as gravitational scattering will generally lead to dynamical heating. A detailed analysis of this effect
is beyond the scope of this work.

In summary, due to the early formation time, the relevant PBH escape velocities are very large in the highly clustered 
scenarios we have studied here, allowing to neglect the effect of GW recoils in our analysis.


  
\section{Conclusions}%
If PBHs are not homogeneously distributed in the Universe but highly clustered, existing bounds on their abundance must be 
re-interpreted. 
 Here we have demonstrated that the resulting limits 
 are not weakened, as claimed previously, but instead strengthened because subsequent merger steps would
 dominate the SGWB. Taking into account constraints from cosmology and direct GW 
 searches, we find that for $\delta_\text{dc,0} > 10^4$ the case of pure PBH DM is firmly excluded in the entire range 
 of initial PBH masses between $10^{-5} \, M_\odot$ and $100 \, M_\odot$. For slightly less conservative assumptions
 about the decrease of $\delta_\text{dc}$ in subsequent merger steps, this even holds for much smaller initial
density contrasts.
 We note that outside this mass range bounds are also very strong~\cite{Carr:2017jsz}, which essentially left this interval
as one out of only two realistic options for explaining all DM in terms of PBHs (the second one arises for much lighter PBHs of $10^{-16} \lesssim m_\text{PBH}/M_\odot \lesssim10^{-11}$ \cite{Niikura:2017zjd}). 

\bigskip
\paragraph*{Note added.---}
After this work was finished,~\cite{Raidal:2018bbj} appeared as a preprint and reopened the question of the stability of PBH binary systems with respect to perturbations by nearby PBHs, which could in particular influence the merger rates entering our results. A detailed analysis of this effect in the context of multiple mergers, which most likely requires $N$-body simulations, is still to be performed.

\bigskip
\vfill 
 \paragraph*{Acknowledgements.---}%
 We thank Thomas Konstandin for useful comments on the manuscript, Florian K\" uhnel and Hardi Veerm\"ae for relevant discussions
 and Cole Miller for encouraging us to spell out why the gravitational wave recoil does not significantly 
 affect our analysis. We would also like to thank the anonymous referees for pertinent comments.
This work is supported by the German Science Foundation (DFG) under the
Collaborative Research Center (SFB) 676 Particles, Strings and the Early Universe as well as the
ERC Starting Grant `NewAve' (638528).
TB wishes to thank McGill university, where part of this manuscript was completed, for support and hospitality.
This research was supported in part by Perimeter Institute for Theoretical Physics. Research at Perimeter 
Institute is supported by the Government of Canada through the Department of Innovation, Science and 
Economic Development and by the Province of Ontario through the Ministry of Research, Innovation and Science.
\newpage

\bibliography{PBHrefs}
 
\end{document}